\documentclass[11pt,letter,notitlepage]{article}

\usepackage[left=1 in, top=1 in, right=1 in, bottom=1 in]{geometry}
\usepackage{graphicx}
\usepackage{amsmath}
\usepackage{amsfonts}
\usepackage{amssymb}
\usepackage{multicol}
\usepackage{fancyhdr}
\usepackage{cases}
\usepackage{sectsty} % needed for paragraph customization
\usepackage{tocloft} % to set spacing for table of contents
\usepackage{hyperref}
\usepackage{url}
\usepackage{authblk}

\usepackage{setspace}

\usepackage{enumitem}
\setlist[itemize]{noitemsep,topsep=0pt}

\usepackage{color}
\definecolor{shiningblue}{rgb}{0.3,0.68,0.89}
\definecolor{brightgreen}{rgb}{0.3,0.8,0.0}

\usepackage[utf8]{inputenc} % Allow Latex-style accents
\usepackage[backend=biber,style=nature]{biblatex}
\parskip = \baselineskip

\subsectionfont{\normalfont\itshape} % customize paragraph font

\newcommand{\dif}[2]{\ensuremath{\frac{d#1}{d#2}}}

\newcommand{\pd}[2]{\ensuremath{\frac{\partial#1}{\partial#2}}}
\newcommand{\pdd}[2]{\ensuremath{\frac{\partial^2#1}{\partial#2^2}}}

\newcommand{\beginsupplement}{
        \setcounter{table}{0}
        \renewcommand{\thetable}{S\arabic{table}}
        \setcounter{figure}{0}
        \renewcommand{\thefigure}{S\arabic{figure}}
        \setcounter{equation}{0}
        \renewcommand{\theequation}{S\arabic{equation}}
} 

\vspace{-3in}

\title{\Large \textbf{Genetic drift in range expansions is very sensitive to density feedback in dispersal and growth}}

\author[1]{Gabriel Birzu\thanks{\nolinkurl{gbirzu@bu.edu}}}
\author[1]{Sakib Matin}
\author[2]{Oskar Hallatschek}
\author[3]{Kirill S. Korolev\thanks{\nolinkurl{korolev@bu.edu}}}
\affil[1]{Department of Physics, Boston University, Boston, MA 02215, USA}
\affil[2]{Departments of Physics and Integrative Biology,University of California, Berkeley, California 94720, USA}
\affil[3]{Department of Physics and Graduate Program in Bioinformatics, Boston University, Boston, MA 02215, USA}

\date{}

\begin{document}
\maketitle

%*************************************************************************
% Abstract
%*************************************************************************
\begin{abstract}
\noindent\textbf{Abstract: }\\
Theory predicts rapid genetic drift in expanding populations due to the serial founder effect at the expansion front. Yet, many natural populations maintain high genetic diversity in the newly colonized regions. Here, we investigate whether density-dependent dispersal could provide a resolution of this paradox. We find that genetic drift is dramatically suppressed when dispersal rates increase with the population density because many more migrants from the diverse, high-density regions arrive at the expansion edge. When density-dependence is weak or negative, the effective population size of the front scales only logarithmically with the carrying capacity. The dependence, however, switches to a sublinear power law and then to a linear increase as the density-dependence becomes strongly positive. To understand these results, we introduce a unified framework that predicts how the strength of genetic drift depends on the density-dependence in both dispersal and growth. This theory reveals that the transitions between different regimes of diversity loss are controlled by a single, universal parameter: the ratio of the expansion velocity to the geometric mean of dispersal and growth rates at expansion edge. Importantly, our results suggest that positive density-dependence could dramatically alter evolution in expanding populations even when its contributions to the expansion velocity is small.
\end{abstract}

\section*{Introduction}

Range expansions and range shifts have become a central theme in population biology because they offer unique opportunities to study rapid adaptation and eco-evolutionary feedbacks~\cite{facon:eco-evolutionary_framework, phillips:life_history_evolution, hellberg:expansion_evolution, dlugosch:devil}. Range expansions are also of great practical interest because they describe the spread of pathogens~\cite{brockmann:network_geometry}, disease vectors~\cite{pastor-satorras:epidemics}, and agricultural pests~\cite{pimentel:invasion_costs}. Predicting and controlling the outcome of range expansions requires a firm understanding of their evolutionary dynamics. Indeed, expansion velocities could increase more than five-fold due to the evolution of faster dispersal~\cite{phillips:toad_acceleration} or decrease due to the rapid accumulation of deleterious mutations at the expansion edge~\cite{bosshard:deleterious_mutations}.

The latter scenario is caused by repeated bottlenecks at the expansion frontier\textemdash a phenomenon known as the founder effect~\cite{slatkin:founder_effect, excoffier:expansion_consequences}. Genetic drift due to the founder effect controls not only the efficacy of natural selection, but also the amount of genetic diversity on which selection can act. In particular, large reductions in diversity are known to negatively impact the resilience of species to environmental stress~\cite{hughes:diversity_review, saccheri:inbreeding}, resistance to parasites~\cite{hughes:diversity_ants}, and adaptation to climate change~\cite{sexton:plasticity, reusch:ecosystem_recovery, jump:option_value}. Thus, the founder effect is a key determinant of the evolutionary potential of an expanding population.

The strength of the founder effect is often inferred from the reduction in neutral genetic diversity at the range margin compared to the population core~\cite{roman:diluting_founder}. In some case, this reduction is dramatic, and a single clone takes over the expansion front~\cite{hollingsworth:clonal_growth}. 
Most empirical studies, however, report only a marginal reduction in the genetic diversity of 10-30 \%~\cite{dlugosch:founding_events}. 
These observations are inconsistent with many theoretical studies and the naive expectation of low effective population size due to a small number of propagules at the leading edge~\cite{slatkin:founder_effect, hallatschek:diversity_wave, waters:founder}.
A number of mechanisms have been proposed to resolve this paradox \cite{frankham:resolving_paradox, schrieber:genetic_paradox_revisited}. For example, diversity loss could be mitigated by multiple introductions~\cite{dlugosch:founding_events} or the bottlenecks could be less sever if the expansion rate is limited by the rate at which suitable habitat becomes available~\cite{hallatschek:constrained_expansions}. The latter scenario could be applicable to range shifts during the climate change or re-expansion of the biosphere following an ice age. While these and other mechanisms could resolve the discrepancy in some specific instances, here we focus on density-dependence as a more general factor that can control the rate of genetic drift at the expansion front.

Density-dependent growth, commonly described as an Allee effect, offers a promising explanation for the high genetic diversity of expanding populations~\cite{waters:founder}.
Previous studies have shown that this influx from the core region reduces the size of the bottleneck and increases the effective population size of the front~\cite{hallatschek:diversity_wave, roques:allee_diversity}. 
Recently, it was found that the effective population size could be proportional to the carrying capacity or scale sublinearly, for example, logarithmically or as a power law~\cite{birzu:fluctuations}. The type of the scaling is very sensitive to the strength of the Allee effect. As a result, effective population size can increase by orders of magnitude in response to a moderate suppression of low-density growth. 
Since growth density-dependence is rarely examined at the expansion front, it is possible that Allee effects are responsible for high genetic diversity in many expanding populations.

Compared to density-dependent growth, density-dependent dispersal received very little attention in the context of diversity loss during expansions. This lack of attention most likely reflects the technical difficulties of studying dispersal both analytically and in the wild, rather than its ecological relevance. 
Indeed, nontrivial density-dependence of dispersal on population density has been documented in many organisms, including plants (\textit{Echinops sphaerocephalus}, \textit{Lythrum alatum}, \textit{Monarda [istulosa})~\cite{levin:bee_mediated_dispersal, antonovics:density_dependence_plants}, arachnids(\textit{Erigone arctica})~\cite{duffey:spiders}, birds (\textit{Parus major}, \textit{Sitta europeae}, \textit{Larus ridibundus})~\cite{matthysen:density_dependence}, and mammals (\textit{Cervus elaphus}, \textit{Capreolus capreolus}, \textit{Marmota olympus})~\cite{matthysen:density_dependence}.
The origins of these strategies are also quite diverse, from cooperative responses to avoid starvation~\cite{tarnita:slime} to mechanisms for population regulation to reduce overcrowding~\cite{duffey:spiders}.
Here, we address this open problem and show that positive density-dependent dispersal can preserve genetic diversity. This result is first established via computer simulations with two neutral alleles invading a patchy one-dimensional landscape. We then formulate a continuum description based on the generalized Fisher-Kolmogorov equation and derive an explicit formula for the effective population size of the front. 
Our analytical treatment differs from previous work which relied on spatial discretization of the population density or the separation of the expansion front into the fully-colonized and newly established populations~\cite{slatkin:founder_effect, peischl:dispersal_evolution}. Therefore, our results could be used to
suggest new ways to study population dynamics in continuous space.

Our theory unifies density-dependence in dispersal and growth and 
shows that both processes can be treated on the same footing, with only a minor increase in complexity.
More importantly, we predict that the scaling of the effective population size with the carrying capacity depends on a single parameter\textemdash the ratio of the expansion velocity to the geometric mean of the growth and dispersal rates at low densities.
In other words, most aspects of density-dependence are fully captured just by the expansion velocity itself. This emergent universality is confirmed by numerical simulations of a diverse set of models with density-dependence in growth and dispersal.

The main implication of our work is that even relatively weak positive density-dependence is sufficient to prevent rapid diversity loss during range expansion. For example, the rate of genetic drift can be two orders of magnitude slower in species that expanded only 10\% faster than they would in the absence of positive feedback in growth or dispersal.
Hence, weak density-dependence could be an overlooked explanation for the surprisingly high diversity of expanding species.

\section*{Methods}

\subsection*{Model}
We simulate range expansions using a variant of the stepping-stone model~\cite{kimura:stepping_stone, korolev:review}. The model consists of a one-dimensional array of islands (demes) with carrying capacity~$N$. To quantify genetic drift, it is sufficient to consider a model with two alleles.
We denote the density of each allele by~$n_i(t,x)$, where~$i \in \{1, 2\}$. 
Each generation, the densities are updated in three steps. In the first step, we compute the densities~$\tilde{n}_i$ after dispersal

\begin{equation}
\label{eq:dispersal_update}
\tilde{n}_i (t, x) = n_i(t,x) - m[n(t, x)] n_i(t, x) + \frac{m[n (t, x-1)]}{2}n_i(t, x-1) + \frac{m[n(t, x + 1)]}{2}n_i(t, x + 1),
\end{equation}

\noindent where~$n(t,x) =  n_1(t,x) + n_2(t,x)$ is the total population density, and~$m(n)$ is the dispersal probability.
Throughout this paper, we focus on parameter regimes which are well described by a continuum model.
In this regime, the assumption that~$m(n)$ only depends on the density in the source deme~$n(t, x)$ is not restrictive because alternative models can be cast in the same form as Eq.~\eqref{eq:dispersal_update}.

In the second step, the expected densities~$\hat{n}_i$ after growth are computed according to

\begin{equation}
\label{eq:growth_update}
\hat{n}_i(t, x) = \left[ 1 + r(\tilde{n}(t, x)) \right] \tilde{n}_i(t, x),
\end{equation}

\noindent where~$r(n)$ is the per capita growth rate. 

The third step accounts for stochasticity in dispersal and growth. This step is performed by drawing~$n_i$ for the next generation from a trinomial distribution

\begin{equation}
\{n_1(t+1), n_2(t+1), 1-n(t+1) \} = \mathrm{Trinomial} \left(N, \frac{\hat{n}_1(t)}{N}, \frac{\hat{n}_2(t)}{N} \right),
\end{equation}

\noindent where~$\mathrm{Trinomial} (M, p_1, p_2)$ is a trinomial distribution with~$M$ tries and success probabilities equal to~$p_1$ and~$p_2$ for alleles~$1$ and~$2$, respectively~\cite{feller:book_vol1}. The above procedure is equivalent to two successive draws from a binomial distribution. The first draw accounts for demographic fluctuations and determines total population density~$n = \mathrm{Bin}(N, \frac{\hat{n}_1 + \hat{n}_2}{N})$. The second draw accounts for genetic drift and determines the genetic composition:~$n_1 = \mathrm{Bin}(N, \frac{\hat{n}_1}{n})$ and~$n_2 = n - n_1$.
This procedure is computationally efficient and enforces a strict carrying capacity in the population bulk.
At the edge of the expansion, where the densities are small, the stochastic dynamics are equivalent to the commonly used Poisson distribution for the number of offspring.

Many studies, both theoretical and experimental, have used linear or power law dependencies to describe density-dependent dispersal~\cite{kawasaki:long_range_taxis, balasuriya:invasion_parameters, golding:rd_models_bacteria}.
To capture a wide range of possible dependencies, we choose the following functional form for the dispersal probability:

\begin{equation}
\label{eq:m_definition}
m(n) = m_0\left[ 1 + A_{b} \left( \frac{n}{N} \right)^{b} \right],
\end{equation}

\noindent where~$m_0$ is a baseline dispersal probability, and~$A_{b}$ quantifies the strength of density-dependence. 
The sign of~$A_b$ controls whether the density-dependence is negative ($A_{b} < 0$) or positive ($A_{b} > 0$). The former is known to enhance unevenness in the spatial distribution of the population and leads to the formation of spatial patterns~\cite{cates:phase_separation, liu:striped_colony}. In contrast, the latter describes strategies that reduce population clustering and helps reduce competition within the population~\cite{kawasaki:long_range_taxis}. 
The exponent~$b$ controls the shape of~$m(n)$. For~$b\ll1$, the function increases sharply at low densities, compared to~$b\gg1$, when most of the increase occurs when~$n$ is close to the carrying capacity (Fig.~\ref{fig:dispersal}a).

Simple logistic growth provides an excellent description of the growth dynamics for many species~\cite{sibly:growth_rate, dennis:density_dependence, barlow:guppies_growth}.
However, this simple model neglects cooperative interactions and other mechanisms that generate an Allee effect~\cite{gandhi:pulled_pushed}. To capture this additional complexity we chose the following widely used functional functional form for the growth rate~\cite{murray:mathematical_biology, aronson:allee_wave}

\begin{equation}
\label{eq:r_definition}
r(n) = r_0 \left (1 - \frac{n}{N}\right) \left (1 + B \frac{n}{N}\right) ,
\end{equation}

\noindent where~$r_0$ is the low-density growth rate, and~$B$ is the strength of the Allee effect.
For~$B \leq 0$ there is no Allee effect. 
For~$B > 0$ we have a weak Allee effect that becomes more pronounced as~$B$ increases. 
We do not consider a strong Allee effect (negative growth at low densities) here since previous work has shown that it lead to the same dynamics as those of Eq.~\eqref{eq:r_definition} with large~$B$~\cite{birzu:fluctuations}. 
In such cases, growth rates at low densities are substantially reduced and multiple introductions take place before the population is established at a new location. As a result, the founder effect is greatly suppressed.

The expansion are performed in a fixed box of size~$L=300$ demes, which moves along with the expansion such that the box is approximately half full (see SI for details). This simulation procedure saves computational resources by excluding demes that are either empty or too far behind the front to affect the genetic composition of the expansion edge. Although this is an approximation, its effect on our results is small if the front width is much less than the box size (see SI). In our case, the typical size of the front width is ten demes, which is an order of magnitude less than~$L$.

\subsection*{Measuring the strength of genetic drift}

\begin{figure}[h!]
\begin{center}
\includegraphics[width=17.8cm]{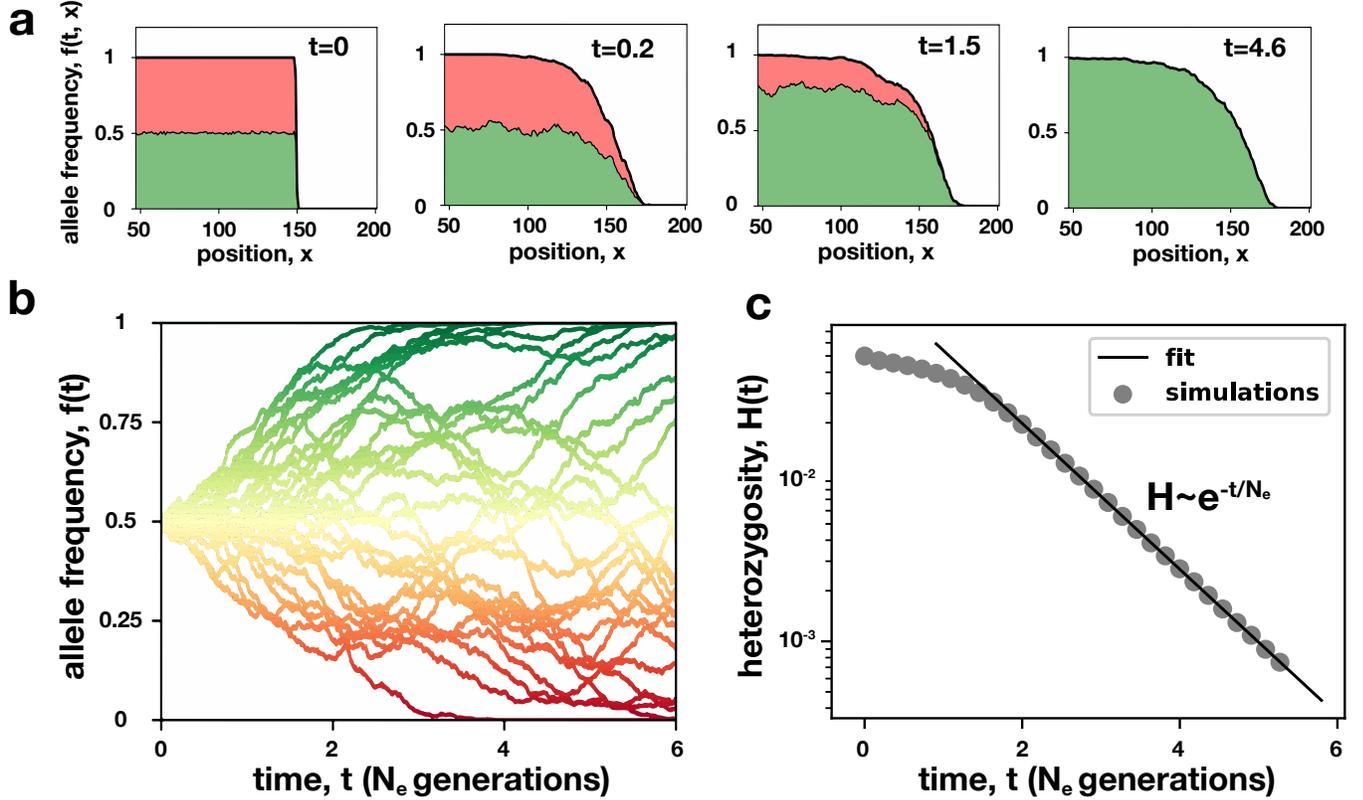}
\caption{\textbf{Genetic drift at the expansion front is described by the average allele frequency.}
~\textbf{(a)} shows how one of the two alleles reaches fixation starting from the initial condition. Note that, at each time point, the spatial variation in the allele frequency is much smaller than the variation between time points. Thus, the composition of the front is well-described by the average fraction of the first allele. The solid line shows the total population density normalized by the carrying capacity. The two colors indicate how this population density is partitioned between the two alleles.
~\textbf{(b)} shows the temporal fluctuations of the average allele frequency~$f(t)$ defined by Eq.~\eqref{eq:f_definition}. The color reflects the proportion of the first allele (green).
~\textbf{(c)} The fluctuations of~$f(t)$ lead to the loss of genetic diversity as seen by the decline of~$H(t)$\textemdash the probability to sample two different alleles; see Eq.~\eqref{eq:H_definition}. After a short transient, the decay of heterozygosity~$H(t)$ is exponential in time, similar to the dynamics of a stationary well-mixed population. Thus, the rate of diversity loss can be quantified by an effective population size~$N_e$. 
Here, $r(n)$ and $D(n)$ are give by Eqs.~\eqref{eq:r_definition} and~\eqref{eq:m_definition} with <list parameters>.}
\label{fig:simulations}
\end{center}  
\end{figure}

We start with a state of maximum diversity by assigning one allele to each individual with equal probability (Fig.~\ref{fig:simulations}a). 
A complete description of this process involves the frequency of the first allele~$f(t,x)=n_1(t,x)/n(t, x)$ that depends on both space and time. 
The spatial variation, however, can be neglected on the long time scale of diversity loss (see Refs.~\cite{hallatschek:diversity_wave} and~\cite{hallatschek:tuned_moments} for derivation). This approximation assumes that the spatial relaxation of~$f(t,x)$ is much faster than the time to fixation, which is valid for realistically large~$N$. The accuracy of replacing~$f(t,x)$ by~$f(t)$ is illustrated in Fig.~\ref{fig:simulations}a, which shows that spatial variation in~$f$ is much smaller than the change in allele frequency which occurs over time.
Thus, we can replace~$f(t,x)$ by the spatial average defined by

\begin{equation}
\label{eq:f_definition}
f(t) = \frac{\sum_{x}n(t, x)f(t, x)}{\sum_{x}n(t, x)}.
\end{equation}

To quantify the rate of diversity loss we introduce the ``heterozygosity''
\begin{equation}
\label{eq:H_definition}
H(t)=\langle f(t)[1 - f(t)]\rangle,
\end{equation}

\noindent where~$\langle \cdot \rangle$ is the ensemble average over independent simulations. 
For populations of haploid asexuals considered here, the heterozygosity represents the probability of randomly choosing two different alleles from the population
After a short transient,~$H(t)$ decays exponentially in time~\cite{hallatschek:diversity_wave, birzu:fluctuations}

\begin{equation}
\label{eq:H_decay}
H(t) \sim e^{-t/N_e}.
\end{equation}

We refer to the timescale of this decay as the effective population size~$N_e$ by drawing an analogy with the dynamics in a well-mixed population. This analogy is known to be imperfect~\cite{charlesworth:ne, sjodin:meaning_Ne, der:generalized_models, rice:multiple_mergers} because some aspects of population dynamics could be distinct from the Wright-Fisher model. Nevertheless,~$N_e$ serves as a convenient and intuitive measure for the rate of genetic drift at the front.

\section*{Results}

\subsection*{Positive density-dependent dispersal suppresses the founder effect}

\begin{figure}[h!]
\begin{center}
\includegraphics[width=8.7cm]{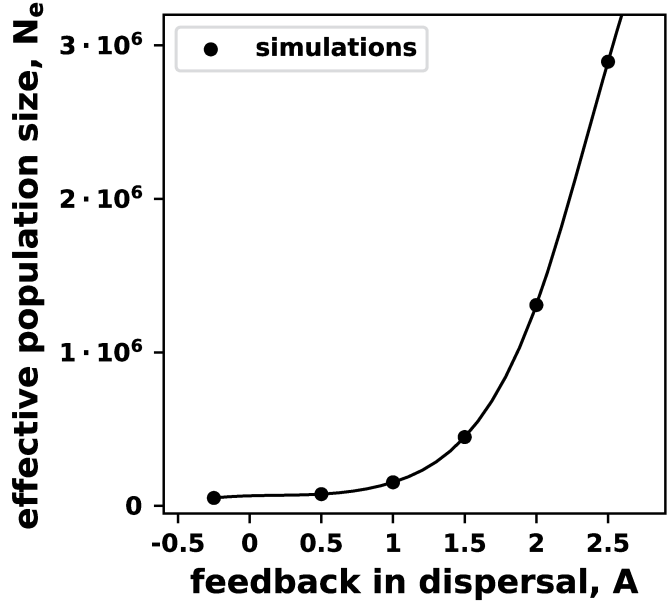}
\caption{\textbf{Diversity loss can be dramatically reduced by positive density-dependent dispersal.}  
The dots represent the effective population size inferred from stochastic simulations with different strength of the density-dependence in dispersal and averaged over 1000 independent runs. The line is cubic spline which we show as a guide to the eye. 
Here, $r(n)$ and $m(n)$ are give by Eqs.~\eqref{eq:r_definition} and~\eqref{eq:m_definition}, with~$N=10^6$,~$b=1$,~$B=0$,~$m_0=0.05$, and~$r_0=0.001$. }
\label{fig:cooperation}
\end{center}  
\end{figure}

How does density-dependent dispersal affect genetic drift in an expanding population?
To answer this question, we simulated range expansions with~$m(n) = m_0(1 + An/N)$ and determined~$N_e$ as a function of~$A$ (see Methods).
We found that while negative density-dependence has a negligible effect on~$N_e$, the effect of positive density-dependence is quite dramatic.
Indeed, increasing~$A$ from~$1$ to~$2.5$ resulted in~$N_e$ changing by an order of magnitude (Fig.~$\ref{fig:cooperation}$).
Moreover, the rate of change in the effective population size increases as the (density) feedback becomes stronger.

These results show that the strength of the founder effect can be influenced just as much by dispersal as by an Allee effect.
What is the connection between these two mechanisms and how dispersal affects diversity loss is what we aim to uncover next.
In order to gain a deeper insight into these processes we extend the theoretical framework used in previous studies~\cite{hallatschek:diversity_wave, birzu:fluctuations} to the case of density-dependent dispersal.

\subsection*{Continuum limit}

Because of the discrete nature of our simulations, an analytical treatment of the dynamics of the heterozygosity is difficult. However, we can make progress by taking the continuum limit of our model, which is a valid approximation when~$r_0 \ll 1$ and~$\sqrt{m_0/r_0} \gg 1$.
In the SI, we derive the continuum equations and find

\begin{equation}
\pd{n_i}{t} = \pdd{}{x} \left[ D(n)n_i \right] + r(n)n_i + \sqrt{\gamma_n(n)n_i} \, \eta_i(t, x),
\label{eq:n_i}
\end{equation}

\noindent where~$\eta_i(t, x)$ is a Gaussian white noise with unit variance,~$n_i(t,x)$ is the density of allele~$i$ at time~$t$ and position~$x$~(which are now continuous variables).

The growth term~$r(n)$ is the same as in Eq.~\eqref{eq:r_definition} in the discrete model apart from an extra factor of~$\Delta t^{-1}$, where~$\Delta t$ is the generation time. The dispersal and noise terms in Eq.~\eqref{eq:n_i} are given by

\begin{subequations} \label{eq:continuum_limit}
\begin{align}
D(n) &= \frac{m(n)\Delta x^2}{2\Delta t}, \label{eq:continuum_limit_a}\\
\gamma_n(n) &= \frac{1}{\Delta t} \left( 1 - \frac{n}{N} \right), \label{eq:continuum_limit_b}
\end{align}
\end{subequations}

\noindent where~$\Delta x$ is the distance between neighboring demes.
The placement of the dispersal function~$D(n)$ inside both derivatives emerges naturally from our microscopic model, but it is not an extra assumption in our model.
The alternative formulation, in which the first term in Eq.~\eqref{eq:n_i} would be replaced by~$\pd{}{x}[D(n) \pd{n}{x}]$, is equivalent to change in the definition of~$D(n)$ (see Ref.~\cite{mendez:aggregation_patterns} and SI).

\subsection*{Deterministic dynamics: pulled and pushed fronts}

Reaction-diffusion fronts of the type given by Eq.~\eqref{eq:n_i} have been studied extensively in the deterministic limit, when noise is absent. In this limit, the total population density obeys the following equation:

\begin{equation}
\label{eq:rd_n}
\pd{n}{t} = \pdd{}{x} \left[ D(n)n \right] + r(n)n.
\end{equation}

Depending on the expansion velocity, these fronts fall into two classes: pulled and pushed waves. The expansion velocity of pulled waves is given by the classic Fisher formula~$v_F=2\sqrt{D(0)r(0)}$, which can be derived by linearizing Eq.~\eqref{eq:rd_n} for small~$n$. Since the velocity of the low density edge of the front determined the expansion velocity, these fronts are said to be ``pulled" by the leading edge. Conversely, pushed expand with~$v>v_F$, as if they are ``pushed" by the nonlinearities in the bulk.
Most investigations into these expansions have focused on~$D(n)=\mathrm{const}$, and have examined how feedback in~$r(n)$ affects the velocity~\cite{aronson:allee_wave, brunet:velocity_cutoff, barton:bistable_dynamics, kessler:velocity_cutoff, murray:mathematical_biology}. 
However, it has been suggested that the transition between pulled and pushed fronts could occur due to either growth or dispersal feedback~\cite{saarloos:review}. Several recent studies corroborate this conclusion~\cite{kawasaki:long_range_taxis, haond:carrying_capacities}. 
In particular, exact analytical solution for~$v$ was obtained in Ref.~\cite{kawasaki:long_range_taxis} for a linear~$D(n)$, which showed the existence of a transition between pulled and pushed waves.
A more extensive discussion of the differences between pulled and pushed fronts, with an emphasis on density-dependent dispersal, can be found in the SI.

\subsection*{Stochastic dynamics: Effective population size}

Most of the previous work on genetic drift in expanding populations has focused on pulled fronts~\cite{brunet:selection_ancestry, hallatschek:tuned_model, good:distribution}. These studies have shown that pulled expansions exhibit large fluctuations in allele frequencies, leading to small effective population sizes, which scale as~$N_e \sim \log^3{N}$~\cite{brunet:selection_ancestry, hallatschek:tuned_model}. 
Recent work extended these results to pushed waves and identified two distinct subclasses of pushed waves, with different scaling of~$N_e$ with~$N$~\cite{birzu:fluctuations}.
However, these studies only considered the case of constant dispersal rates. Here, we derive a complete theory of genetic drift during range expansions that accounts for arbitrary density-dependence in dispersal and growth.

In the SI, we generalize the calculations from Ref.~\cite{birzu:fluctuations} to density-dependent dispersal. The calculation relies on treating the front deterministically up to an appropriately chosen cutoff at low densities. To lowest order in~$N^{-1}$, the effective population size is given by

\begin{equation}
N_e = \left\{
\begin{aligned}
& C_1 \ln^3{N}, \quad\quad \nu = 1,\\
& C_2 N^{\frac{2\sqrt{1 - 1/\nu^2}}{1 - \sqrt{1 - 1/\nu^2}}}, \quad\quad 1 < \nu < \frac{3}{2\sqrt{2}},\\
& C_3 N, \quad\quad \nu \ge \frac{3}{2\sqrt{2}},\\
\end{aligned}
\right.
\label{eq:Ne}
\end{equation}

\noindent where we have defined the normalized expansion velocity

\begin{equation}
\nu = \frac{v}{2\sqrt{r(0)D(0)}}.
\label{eq:nu}
\end{equation}

\begin{figure}[ht!]
\begin{center}
\includegraphics[width=17.8cm]{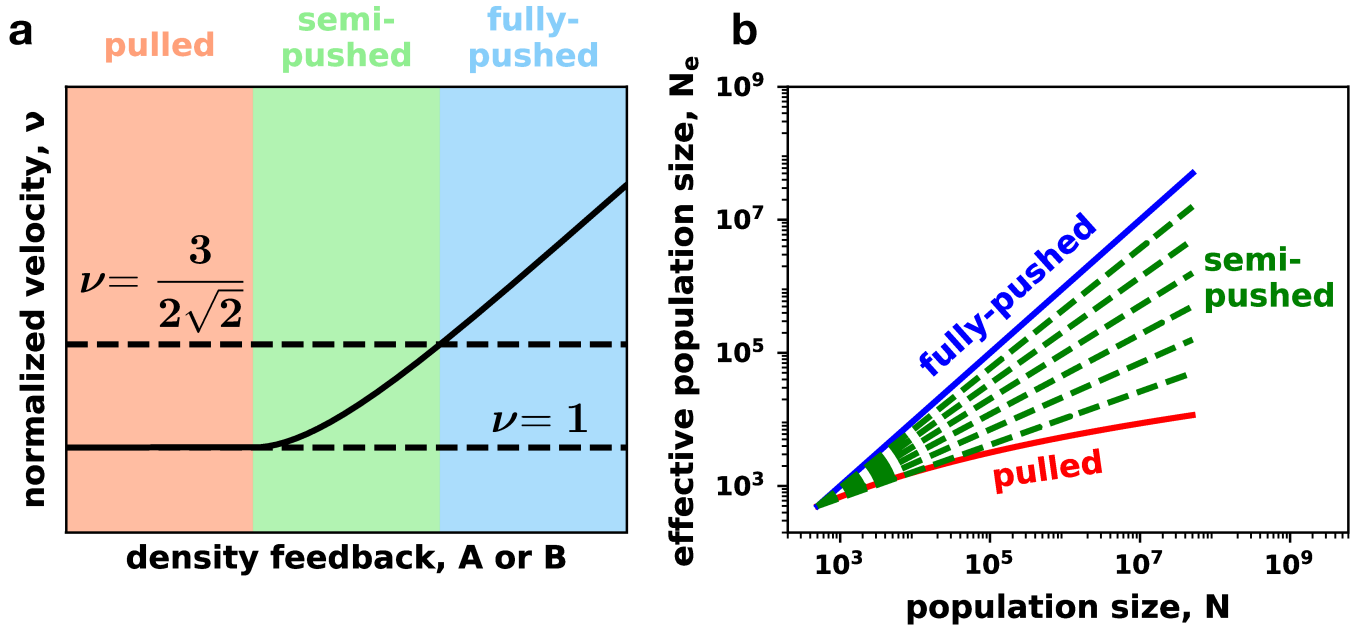}
\caption{\textbf{Theory predicts three distinct regimes of diversity loss.}
~\textbf{(a)} For negative or weakly-positive feedback, the expansions are pulled, and their velocity depends only on the growth and dispersal at low densities. Thus, the normalized velocity,~$\nu$ strictly equals one for a range of density feedback. The transition to pushed waves is marked by~$\nu>1$, which increases with the strength of the feedback in this regime. Our theory predicts a second transition at~$\nu=\frac{3}{2\sqrt{2}}$ that separates semi-pushed from fully-pushed expansions. Depending on the value of~$\nu$, expansions are classified as either pulled ($\nu = 1$, red-shaded region), semi-pushed ($1 < \nu < \frac{3}{2\sqrt{2}}$, green-shaded region), or fully-pushed ($\nu > \frac{3}{2\sqrt{2}}$, blue-shaded region). 
The differences between these three wave classes are illustrated in~\textbf{(b)}, which shows the dependence of the effective population size on the carrying capacity predicted by Eq.~\eqref{eq:Ne}. For fully-pushed waves, $Ne$ increases linearly with~$N$ (slope one on the log-log plot). For semi-pushed waves, $Ne\sim N^\alpha$ with~$\alpha\in(0,1)$ that depends only on~$\nu$. For pulled waves,~$Ne\sim \ln^3N$. 
In (a), we show the exact solution of Eq.~\eqref{eq:rd_n} for~$D(n)=D_0(1 + An/N)$ and~$r(n)=r_0(1-n/N)$ from Ref.~\cite{kawasaki:long_range_taxis}. In (b), we sketch the asymptotic scaling predicted by Eq.~\eqref{eq:Ne} by choosing different values of~$C_1$,~$C_2$,~$C_3$.}
\label{fig:theory}
\end{center}  
\end{figure}

An intuitive interpretation of~$\nu$ is the ratio between the actual velocity of the expansion~$v$ and the velocity inferred from Fisher's formula. Clearly, pulled waves have~$\nu=1$, while pushed waves have~$\nu>1$.
The constants~$C_1$,~$C_2$, and~$C_3$ in Eq.~\eqref{eq:Ne} are independent of~$N$ and are given in the SI.

Two critical values of~$\nu$ divide expansions into three regimes (Fig.~\ref{fig:theory}a). For low values of the feedback, the expansions are pulled and the logarithmic scaling of~$N_e$ with~$N$ is now well established~\cite{brunet:selection_ancestry, hallatschek:tuned_model, birzu:fluctuations}. 
At higher values of the feedback, we find two types of pushed waves: semi-pushed waves, for which~$N_e$ scales as a sublinear power law in~$N$, and fully-pushed waves, for which we recover the well-mixed scaling~$N_e \sim N$ (Fig.~\ref{fig:theory}b).

The surprising implication of Eq.~\eqref{eq:Ne} is that the type of the expansion and the scaling of~$N_e$ with~$N$ depend on a single parameter, the normalized velocity~$\nu$. In particular, it is of no consequence whether high values of~$\nu$ arise due to nonlinear dispersal or an Allee effect. Furthermore, all complexity of~$D(n)$ and~$r(n)$ are fully encoded in~$\nu$.

\subsection*{Testing universality}

We performed two tests of the universal behavior of~$N_e(N,\nu)$.
First, we checked that different dispersal models lead to the same scaling and that the scaling exponent matches the prediction from Eq.~\eqref{eq:Ne}.
Second, we confirmed that the scaling is unchanged when both the density-dependence in dispersal and the strength of an Allee effect are combined in different proportions to keep~$\nu$ constant.

\begin{figure}[h!]
\begin{center}
\includegraphics[width=17.8cm]{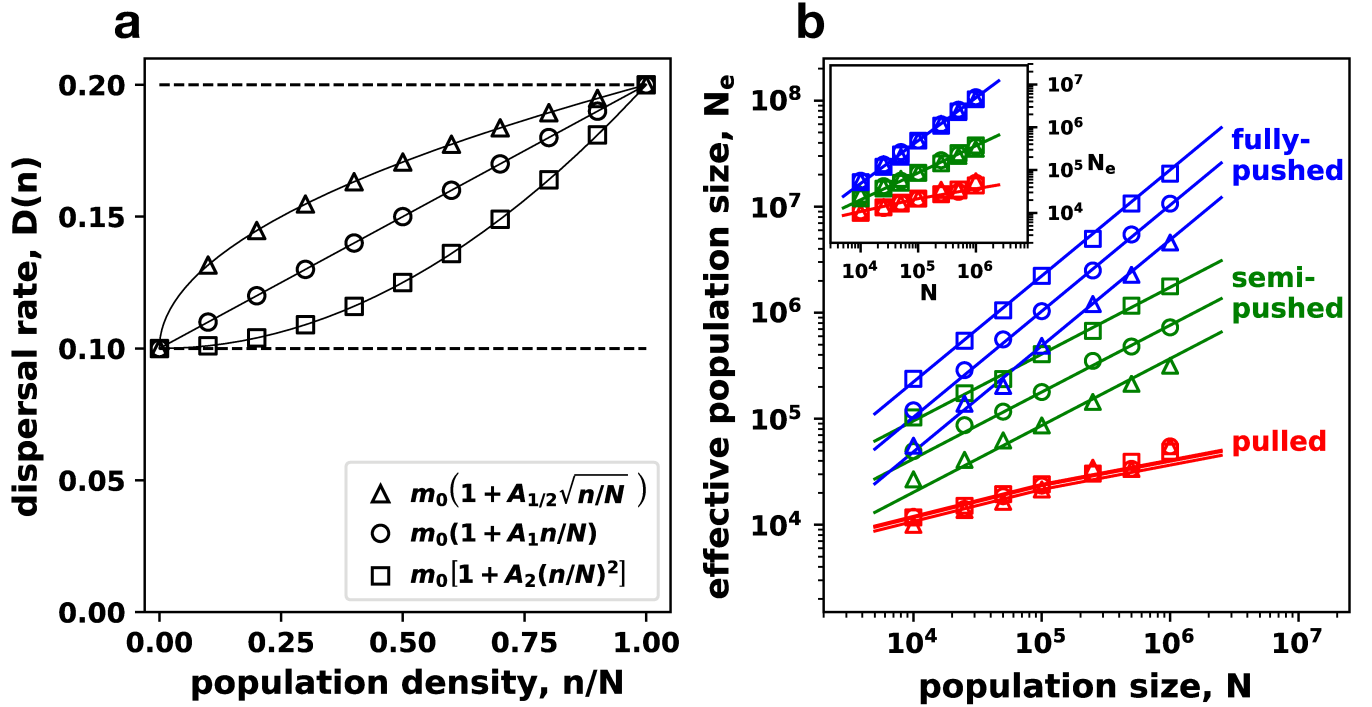}
\caption{\textbf{Simulations of different dispersal models confirm theoretical predictions for~$N_e(N)$.}~
\textbf{(a)} illustrates~$D(n)$ for three models with different exponent~$b$, which controls how quickly dispersal increases with the population density; see Eq.~\eqref{eq:m_definition}.
~\textbf{(b)} shows~$N_e(N)$ obtained from computer simulations. The symbols are the same as in (a) and indicate the dispersal model. The lines are theoretical predictions. The exponents are computed from Eq.~\eqref{eq:Ne} and the unknown proportionality constants are adjusted for the best fit. For visual clarity, we chose the model parameters such that all three semi-pushed waves have the same~$\nu$, so the green lines are parallel to each other. The agreement between the theory and the three models is shown more clearly in the inset where~$Ne$ is normalized by the fitting constant~$C_i$ from Eq.~\eqref{eq:Ne}. The collapse of the data confirms that the scaling exponent depends only on~$\nu$ and not on the details of the dispersal model. 
Stochastic  simulations were done using~$m_0=0.05$ and~$r_0=0.001$, and averaged over~$1000$ runs.}
\label{fig:dispersal}
\end{center}  
\end{figure}

To carry out the first test we measure~$N_e(N)$ in simulations with~$m(n)$ given by Eq.~\eqref{eq:m_definition} for three different values of the exponent~$b$~(see Fig.~\ref{fig:dispersal}a).
For each model, we choose the feedback~$A_b$ such that the normalized velocities of the expansions are equal in each of the three regimes.

For pulled waves, the dynamics are fully captured by the linearized equation so~$N_e$ should be independent of the nonlinearities in the dispersal model. Indeed, Fig.~\ref{fig:dispersal}b shows that the results for pulled waves collapse on the same curve, which is well fit by~$\log^3N$.
For semi-pushed waves, we find a quantitative agreement between the simulations and the power law scaling predicted by our theory (see SI). For fully-pushed waves, the three models exhibit linear scaling, as expected. Thus, our simulations confirm the predicted functional dependence of~$N_e$ on~$N$ in all three regimes, independent of the shape of~$m(n)$.

\begin{figure}[ht!]
\begin{center}
\includegraphics[width=17.8cm]{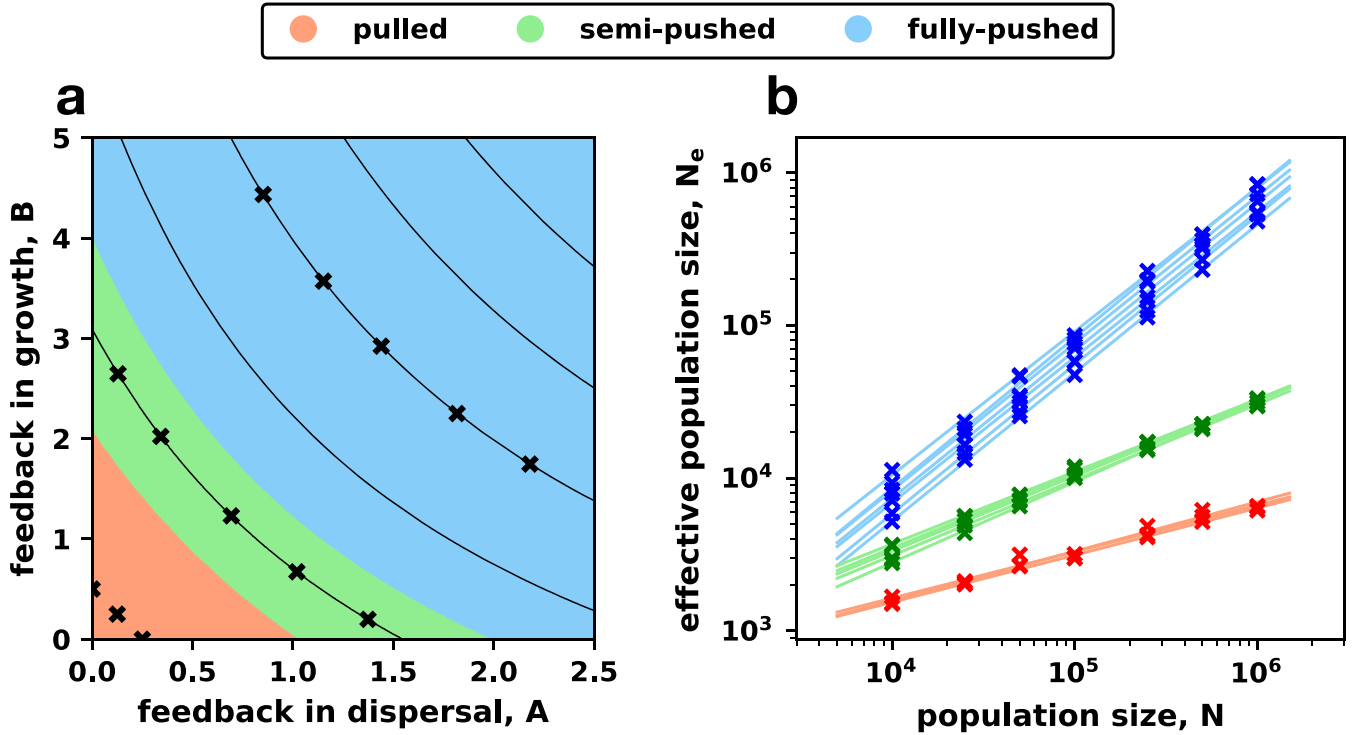}
\caption{\textbf{Diversity loss is insensitive to the source of the positive-density dependence.}
~\textbf{(a)} illustrates how positive feedback in dispersal and growth affect expansion velocity. Lines of constant velocity are shown with thin solid lines. Note that all pulled waves have the same velocity. The crosses mark the parameters chosen for the simulations in panel~(b). The different expansion classes are indicated with color. 
~\textbf{(b)} shows~$N_e(N)$ for different values of~$A$ and~$B$ taken along velocity isoclines in each of the three wave classes~(crosses in a). The thin lines are the best linear fits to the data. Within each class, the lines are parallel to each other confirming that the scaling exponents depend on density-feedback only through~$\nu$ and not through~$A$ and~$B$ separately.
The velocities in (a) were obtained by numerically integrating Eq.~\eqref{eq:rd_n} with~$D(n)$ given by Eq.~\eqref{eq:m_definition} and~$r(n)$ by Eq.~\eqref{eq:r_definition}.
The crosses in (a) are equidistantly chosen on the isoclines~$\nu=1$,~$\nu=\frac{1}{2} + \frac{3}{4\sqrt{2}}$, and~$\nu=1.3$.
}
\label{fig:universality}
\end{center}  
\end{figure}

To carry out the second test of our theory, we vary both the growth and dispersal parameters, while keeping~$\nu$ constant, for~$m(n)$ given by Eq.~\eqref{eq:m_definition} and~$r(n)$ given by Eq.~\eqref{eq:r_definition}. 
This is conceptually illustrated by Fig.~\ref{fig:universality}a.
According to our theory, such changes in parameters should not affect the scaling~$N_e$ with~$N$. 
This expectation is confirmed by our simulations in Fig.~\ref{fig:universality}b.
Similar to the previous test, all pulled waves collapse to the same curve.
In contrast, for semi-pushed and fully-pushed waves only the scaling is predicted to be universal. Thus the intercepts in Fig.~\ref{fig:universality}b are given the nonuniversal constants~$C_2$ and~$C_3$.
Across the isocline, the change in~$C_2$ is small and the curves nearly collapse, while~$C_3$ changes by a factor of~$2$. Still, this change is much smaller than the change in~$N_e$ due to~$N$ or~$\nu$, which is of several orders of magnitude.

\section*{Discussion}

Density-dependent dispersal has been found in empirical studies across the natural world\textemdash from plants and animals~\cite{antonovics:density_dependence_plants, matthysen:density_dependence} to insects and microbes~\cite{duffey:spiders, levin:bee_mediated_dispersal, kearns:bacterial_swarming, vanDitmarsch:convergent_evolution}.
Both theoretical and empirical studies have shown that optimal dispersal strategy can offer significant selective advantage, or and even ensure the ultimate survival of the species~\cite{peischl:dispersal_evolution, travis:dispersal_evolution, kawasaki:long_range_taxis, dahmen:windy_oasis}. 
While the evolution of dispersal strategies has been extensively studied, the reverse problem\textemdash the effect of dispersal on evolution\textemdash has largely been unexplored. 
Here we have addressed this question in the context of range expansions.

We found that positive density-dependent dispersal dramatically reduces the rate of diversity loss, while negative density-dependence has little effect.
The intuition behind this result is that the flux of immigrants from the population core become comparable to the flux of early colonizers.
To explain this behavior quantitatively, we computed~$N_e$ for a general model with arbitrary density-dependence in dispersal and growth.
Using this theory, we showed that the change in the effective population size is caused by two transitions, both marked by changes in the scaling of~$N_e$ with the carrying capacity~$N$.

Weak or negative density-dependence results in pulled expansions with~$\nu=1$ and~$N_e\sim\ln^3N$. Expansions become semi-pushed as the density dependence increases, and~$\nu$ begins to exceed~$1$. Although semi-pushed waves rely on dispersal and growth behind the expansion edge, their population dynamics are still highly stochastic, and~$N_e$ grows as a sublinear power law of~$N$. Further increase in the strength of the density dependence leads to a second transition from semi-pushed to fully-pushed waves~($\nu>\frac{3}{2\sqrt{2}}>1$). The fluctuations in fully-pushed waves are suppressed, and~$N_e\sim N$. Thus, the naive expectation that the effective population size is given by the number of organisms at the front is only valid for range expansions with strong positive density dependence. Since the strength of the density-dependence is fully encoded in~$\nu$, one can determine the class of the range expansion by comparing the observed expansion velocity to the Fisher velocity expected from the growth and dispersal rates at low population densities~\cite{fisher:wave, kolmogorov:wave, skellam:wave}.

Many of our results parallel those for density-independent dispersal in Ref.~\cite{birzu:fluctuations}. However, it is a highly nontrivial result of our work that density-dependent dispersal and growth have similar effects on population dynamics and can be treated in a unified way. In particular, the functional dependence of~$N_e$ on~$N$ and many other ecological and evolutionary processes in an expanding population are controlled by a single parameter, the normalized velocity $\nu$. To determine~$\nu$, one needs to measure only three quantities: the expansion velocity~$v$, the dispersal rate at low densities~$D(0)$, and the growth rate at low densities~$r(0)$. 
Once~$\nu$ known it can be used to predict the fixation probabilities of neutral mutations or the amount genetic diversity in the population.
Even more importantly,~$\nu$ can be used to compare different species or different environments. Such standardization would allow researchers to separate the intrinsic dynamics of the population from their specific biological characteristics such as the genetic architecture or mating system.

Our results suggest an enticing although yet untested possibility of different genealogies in pulled, semi-pushed, and fully-pushed expansions. In well-mixed populations, distinct scaling of~$N_e$ with~$N$ is known to be a manifestation of fundamentally different structure of the genealogical trees. The~$N_e\sim\ln^3N$ scaling in pulled waves is identical to that in large populations evolving due to weak, but numerous beneficial mutations. The genealogies in such rapidly adapting populations are described by the Bolthausen-Sznitman coalescent with multiple mergers~\cite{bolthausen:coalescent}. The standard~$N_e\sim N$ scaling is observed, for example, for strictly neutral evolution, which leads to the Kingman coalescent for ancestral lineages~\cite{kingman:genealogies}. These two analogies suggest that the genealogies of expanding populations are described by a family of~$\Lambda$-coalescents that has Bolthausen-Sznitman and Kingman coalescents as two limiting cases for pulled and fully-pushed waves respectively. If this conjecture is correct, we can gain a better understanding of the eco-evolutionary dynamics of range expansions by analyzing the properties of genealogical trees that go beyond the rate of diversity loss.

In addition to genetics, other properties of the front could be controlled by density feedback.
and calculate the effective diffusion constant of the expansion front,~$D_f$. Similar to previous results~\cite{brunet:selection_ancestry, birzu:fluctuations}, we find that~$D_f^{-1}$ scales with~$N$ in the same way as~$N_e$. Thus, the behavior of both evolutionary~($N_e$) and ecological~($D_f$) properties of the expansion falls into one of the three distinct classes, depending on the value of the normalized velocity~$\nu$.

The linear scaling of~$N_e$ with the population size we observe in fully-pushed waves is apparently very similar to the main result of Ref.~\cite{slatkin:founder_effect}. However, in deriving this result, Slatkin et al. relied on approximating the propagation by successive founder events, in which the population at the edge of the expansion reaches carrying capacity before the next deme is colonized. While this approximation is frequently made in the literature~\cite{peischl:deleterious_accumulation, peischl:load, phillips:sorting, peischl:dispersal_evolution}, our results highlight an important limitation of this approach. The approximation effectively divides the front into fully-colonized and newly established populations, and reduces the evolutionary dynamics to those of a two deme model. The loss of diversity in the population can then either be limited by migration\textemdash and hence independent of~$N$\textemdash or by the time for diversity loss within one deme\textemdash which is proportional to~$N$. The nontrivial scaling we observe for pulled and semi-pushed waves is a consequence of the broad distribution of fixation probabilities across the expansion front (see SI from Ref.~\cite{birzu:fluctuations}). Thus, to capture the correct behavior of these waves, it is necessary to consider the full dynamics across the width of the front.

How populations maintain genetic diversity during expansions\textemdash also known as the genetic paradox of invasions\textemdash has been a major question in evolutionary ecology~\cite{perez:genetic_adaptation_paradox}. 
This paradox hinges on the studies of pulled expansions, which indeed have a small effective population size that grows only logarithmically with the number of organisms at the expansion front. For fully-pushed expansions, however, the paradox disappears completely because the effective population size is comparable to the census population size of the front, and the rate of diversity loss is therefore exceedingly slow. In this paper, we have shown that pushed expansions and weak genetic drift are quite generic in the presence of positive feedback in growth and dispersal. The prevalence of pulled, semi-pushed, and fully-pushed expansions in the wild is an important and fascinating topic for future investigations. However,
given the prevalence of density-dependence in both dispersal and growth reported in field studies~\cite{matthysen:density_dependence, kramer:allee_evidence}, it is possible that a strong founder effect is more of an exception rather than a rule.

%**************************************************************************
% Acknowledgements
%**************************************************************************
\textbf{\large {Acknowledgements}}\\
This work was supported by Simons Foundation Grants \#409704 (to K.S.K.) and \#327934 (to O.H.); by the Research Corporation for Science Advancement through Cottrell Scholar Award \#24010 (to K.S.K.) and the Scialog grant \#26119 (to K.S.K.); by the Moore foundation grant \#6790.08 (to K.S.K.); by the National Science Foundation Career Award \#1555330 (to O. H.); and by the the National Institute of General Medical Sciences of the National Institutes of Health Award R01GM115851 (to O. H.). Simulations were carried out on the Boston University Shared Computing Cluster.

%*************************************************************************
% Supplemental Information
%*************************************************************************
\clearpage
\beginsupplement

\title{\Large Supplemental Information}
\date{}
\maketitle

This Supplemental Information (SI) contains an expanded description of our computer simulations and data analysis, and detailed derivations of the results from the main text. In addition, exact analytical results are derived for some models of density-dependence.

\tableofcontents

\section{Density-dependence and positive feedback in models of range expansions} \label{sec:dispersal}

We model population expansions using a costless, short-ranged dispersal in the continuum limit. These three assumptions imply that the change in the population density due to dispersal is given by the divergence of a flux:

\begin{equation}
\label{eq:change_flux}
\pd{n}{t} = -\pd{}{x}\left[ J\left(n, \pd{n}{x} \right) \right]
\end{equation}

The flux~$J$ can be written in two equivalent forms:

\begin{equation}
\label{eq:flux_hermitian}
J = -D(n)\pd{n}{x},
\end{equation}

or

\begin{equation}
\label{eq:flux_tilde}
J = -\pd{}{x} \left[ \tilde{D}(n) n \right].
\end{equation}

Both~$D(n)$ and~$\tilde{D}(n)$ can be interpreted as effective diffusion constants and are related by

\begin{equation}
\label{eq:D_Dtilde}
D(n) = \tilde{D}(n) + \tilde{D}'(n)n.
\end{equation}

In our analytical derivation, we use~$D(n)$ because this form of dispersal can be more easily converted to a Hermitian operator (see Sec.~\ref{sec:Ne}). In our simulations, we use the other form because dispersal that only depends on the population density at the source deme is naturally described by~$\tilde{D}(n) = m(n)/2$. Note that we used the simpler notation~$D(n)$ in the main text for the model given by Eq.~\eqref{eq:flux_tilde}. However, throughout the SI, we follow the convention given by Eqs.~\eqref{eq:flux_tilde} and~\eqref{eq:flux_hermitian}.

\section{Traveling wave solutions: pulled and pushed expansions} \label{sec:pulled_pushed}

In this section, we explain the difference between pulled and pushed waves emphasizing the role of dispersal.
Prior work on this topic largely assumes density-independent dispersal and considers how the fronts change due to nonlinear growth.
Most of the results, however, can be extended to density-dependent dispersal.
Below, we only provide a minimal, heuristic discussion based on the approach from
Ref.~\cite{birzu:fluctuations} (SI, Sec. II).
A more thorough presentation of these ideas can be found in Ref.~\cite{saarloos:review}.

The equation we consider in the main text is

\begin{equation}
\frac{\partial n}{\partial t} = \pd{}{x} \left[ D(n) \pd{n}{x} \right] + r(n) n.
\label{eq:deterministic_n}
\end{equation}

Assuming sufficiently localized initial conditions, this equation admits a traveling wave solution of the following form:

\begin{equation}
\label{eq:comoving}
\begin{aligned}
&n(t,x) = n(\zeta), \\ 
&\zeta = x - vt, \\
\end{aligned}
\end{equation}

\noindent where without loss of generality we focus on the right-moving part of the expansion.

Upon substituting Eq.~\eqref{eq:comoving} into Eq.~\eqref{eq:deterministic_n}, we obtain the equation for $v$ and~$n(\zeta)$:

\begin{equation}
\label{eq:reduced_ode}
[D(n)n']' + vn' + r(n)n = 0,
\end{equation}

\noindent where primes denote derivatives with respect to~$\zeta$. 
The boundary conditions for Eq.~\eqref{eq:reduced_ode} are given by

\begin{equation}
\label{eq:boundary}
\begin{aligned}
&n(-\infty) = N,\\
&n(+\infty) = 0.
\end{aligned}
\end{equation}

The solutions of~Eq.~(\ref{eq:reduced_ode}) clearly have a translational degree of freedom, i.e., if~$n(\zeta)$ is a solution, then~$n(\zeta + \mathrm{const})$ is a solution. We can eliminate this degree of freedom by choosing the references frame such that~$n(0)=N/2$. The general solution of Eq.~(\ref{eq:reduced_ode}) then has only one remaining degree of freedom, which corresponds to the value of~$n'(0)$. For a given~$v$, the existence of the solution depends on whether the value of~$n'(0)$ can be adjusted to match the boundary conditions specified by Eq.~(\ref{eq:boundary}).

Depending on the behavior of the solution at~$\zeta\to\pm\infty$, each boundary condition may or may not provide a constraint on the solution and, therefore, remove either one or zero degrees of freedom. To determine the number of constraints, we linearize Eq.~\eqref{eq:reduced_ode} near each of the boundaries.

\subsection{Behavior near the boundaries}

For~$\zeta\to-\infty$, we let~$u=N-n$ and obtain:

\begin{equation}
D(N)u'' + vu' + r'(N)Nu = 0,\\
\label{eq:reduced_ode_bulk}
\end{equation}

\noindent which has the following solution:

\begin{equation}
\label{eq:solution_bulk}
u = A_b e^{k_b \zeta} + C_b e^{q_b \zeta},
\end{equation}

\noindent where

\begin{equation}
\label{eq:k_q_bulk}
\begin{aligned}
&k_b = \frac{\sqrt{v^2 - 4D(N)r'(N)N} -v}{2D(N)},\\
&q_b = \frac{-\sqrt{v^2 - 4D(N)r'(N)N} - v}{2D(N)}.\\
\end{aligned}
\end{equation}

\noindent Here, we used subscript~$b$ to indicate that we refer to the behavior of the population bulk. Because the carrying capacity is an attractive fixed point,~$dr/dn$ is negative at~$n=N$, while~$D(N) > 0$, and therefore~$k>0$ and~$q<0$. We then conclude that the boundary condition at~$\zeta \to -\infty$ requires that~$C_b=0$ and thus selects a unique value of~$n'(0)$. 

Next, we analyze the behavior at the front and linearize Eq.~\eqref{eq:reduced_ode} for small~$n$:

\begin{equation}
D(0)n'' + vn' + r(0)n = 0,
\label{eq:reduced_ode_front}
\end{equation}

\noindent which has the following solution:

\begin{equation}
\label{eq:solution_front}
n = A e^{-k \zeta} + C e^{-q \zeta},
\end{equation}

\noindent where

\begin{equation}
\label{eq:k_q_front}
\begin{aligned}
&k = \frac{v+\sqrt{v^2 - 4D(0)r(0)}}{2D(0)},\\
&q = \frac{v-\sqrt{v^2 - 4D(0)r(0)}}{2D(0)}.\\
\end{aligned}
\end{equation}

The implications of~Eq.~\eqref{eq:k_q_front} depend on the sign of~$r(0)$. When~$r(0) < 0$ (strong Allee effect),~$q<0$ and, therefore, one needs to set~$C=0$. This constraint uniquely determines the wave velocity. Such waves are classified as pushed because the linearization of Eq.~\eqref{eq:deterministic_n} has only solutions that decay in time.

For~$r(0)>0$, the analysis is more subtle because, naively, the behavior at the front does not fully constrain the velocity of the wave.
Indeed, one can find a solution of Eq.~\eqref{eq:reduced_ode} satisfying Eq.~\eqref{eq:comoving} for arbitrary~$v \ge 2 \sqrt{r(0) D(0)}$.
For pulled waves,~$q=k$ and the minimum velocity is selected by the growth dynamics of the linearization of Eq.~\eqref{eq:deterministic_n}, which can be proved exactly (see Sec. II in the SI of Ref.~\cite{birzu:fluctuations}). In this case, the velocity and front decay rate are given by

\begin{equation}
\label{eq:v_f}
v_{\mathrm{\textsc{f}}} = 2 \sqrt{r(0) D(0)}.
\end{equation}

and

\begin{equation}
k_{\mathrm{\textsc{f}}} = \sqrt{\frac{r(0)}{D(0)}},
\label{eq:k_f}
\end{equation}

To understand the origin of pushed waves, we need to consider the behavior of the population density near the front for~$v>v_{\mathrm{\textsc{f}}}$. 
Equations~\eqref{eq:solution_front} and \eqref{eq:k_q_front} predict that~$n(\zeta)$ is a sum of two exponentially decaying terms with different decay rates: one term decays with the rate~$k>k_{\mathrm{\textsc{f}}}$, but the other with the rate~$q<k_{\mathrm{\textsc{f}}}$.
Note that we use~$v_F=$ and~$k_F$ defined above as a convenient parameter combination, even though they do not describe the velocity or the decay rate of the population density of the pushed front.

Rigorous analysis of the linearized equation for the front shows that~$n(t,x)$ must decay to zero with a rate at least as large as~$k_{\mathrm{\textsc{f}}}$~\cite{saarloos:review}. 
Therefore~$C=0$. This extra requirement makes the problem of satisfying the boundary conditions overdetermined. As a result, there are two alternatives. First, there could be no solutions for any~$v>v_{\mathrm{\textsc{f}}}$. In this case, the expansion must be pulled because it is the only feasible solution. Second, a special value of~$v$ exists for which the boundary conditions at~$\zeta\to\pm\infty$ can be satisfied simultaneously. In this case, the wave is pushed because the pulled expansion at low~$n$ is quickly overtaken by the faster expansion from the bulk.

To summarize, expansions can be either pulled or pushed depending on the relative strength of the growth and dispersal behind vs. at the front. Pushed waves expand with~$v>v_{\mathrm{\textsc{f}}}$, and their profile decays as~$e^{-k\zeta}$. In contrast, pulled waves expand with~$v=v_{\mathrm{\textsc{f}}}$, and their profile decays as~$e^{-k_\mathrm{\textsc{f}}\zeta}$. In general, the class of a wave cannot be determined without solving the full nonlinear problem given by~Eq.~\eqref{eq:reduced_ode}.

%**************************************************************************
% Neutral markers
%**************************************************************************
\section{Dynamics of neutral markers and fixation probabilities} \label{sec:fixation_probability}

Heritable neutral markers provide an easy way to study the changes in genetic diversity that occur during a population expansion. These dynamics can be studied either forward-in-time or backward-in-time. Here, we mostly focus on forward-in-time dynamics. A more complete discussion of the differences between the two approaches can be found in Ref.~\cite{birzu:fluctuations} (SI, Sec. III). The goal of this section is to introduce the notation that we use in the following sections and explain how density-dependent dispersal affects the dynamics of neutral alleles.

\subsection{Deterministic dynamics of a neutral marker}
We consider an expanding population composed of two subpopulations labeled by a neutral genetic marker inherited from parent to offspring. In the following we study two alleles for simplicity, but the results can easily be generalized to an arbitrary number of alleles~\cite{korolev:review, birzu:fluctuations}. Since the growth and dispersal rates are the same for both subpopulations, both~$n_i$ obey the equation

\begin{equation}
\pd{n_i}{t} = \pd{}{x}\left[ D(n) \pd{n_i}{x} \right] + r(n)n_i,
\label{eq:ni_wave}
\end{equation}

\noindent where~$n=\sum_i n_i$ is the total population density. It is convenient to define the relative frequency of one of the markers:~$f = n_1/n$. Taking the time derivative of~$f$ and using~\eqref{eq:deterministic_n} and~\eqref{eq:ni_wave} gives the equation for the relative frequency:

\begin{equation}
\pd{f}{t} = \pd{}{x} \left[ D(n)\pd{f}{x} \right] + 2D(n)\pd{\ln n}{x} \pd{f}{x}.
\label{eq:fi}
\end{equation}

The new advection-like term arrises from the nonlinear change of variables and reflects the larger change in~$f$ due to dispersal from locations with higher population density. The net effect of this term is a ``flow" of~$f$ from the back to the tip of the expansion. It is convenient for further analysis to shift to the comoving reference frame:

\begin{equation}
\pd{f}{t} = \pd{}{\zeta} \left[ D(n)\pd{f}{\zeta} \right] + \left[ v + 2D(n)\pd{\ln n}{\zeta} \right] \pd{f}{\zeta}.
\label{eq:fi_det}
\end{equation}

\subsection{Stochastic dynamics}

The equations derived in the previous subsection apply to deterministic fronts. However, genetic diversity is lost due to genetic drift, an inherently stochastic process. Here, we present the equations for~$n$ and~$f$ taking into account demographic fluctuations and genetic drift. 
Because the fluctuations depend on the details of the reproductive process, we introduce two new functions of the population density:~$\gamma_n(n)$ and~$\gamma_f(n)$, that describe the strength of fluctuations in~$n$ and~$f$. The derivation of the stochastic equations for~$n(t,\zeta)$ and~$f(t,\zeta)$ is analogous to the~$D(n)=\textrm{const}$ case presented in Ref.~\cite{birzu:fluctuations} (SI, Sec. III). Below we give the final result:

\begin{equation}
\pd{n}{t} = \pd{}{\zeta}\left[ D(n) \pd{n}{\zeta} \right] + r(n)n + \sqrt{\gamma_n(n)n} \, \eta(t, \zeta),
\label{eq:n_stochastic}
\end{equation}

\begin{equation}
\pd{f}{t} = \pd{}{\zeta} \left[ D(n)\pd{f}{\zeta} \right] + \left[ v + 2D(n)\pd{\ln n}{\zeta} \right] \pd{f}{\zeta}  + \sqrt{\frac{\gamma_f(n)f(1 - f)}{n}} \, \eta_f(t, \zeta).
\label{eq:fi_stochastic}
\end{equation}

%**************************************************************************
% Effective population size
%**************************************************************************
\section{Effective population size of the front} \label{sec:Ne}

\subsection{Forward-in-time dynamics of the heterozygosity}

We measure the genetic diversity in the expansion front using~$H(t)$\textemdash the heterozygosity averaged over the front. To derive an expression for~$H(t)$, we follow Refs.~\cite{hallatschek:diversity_wave, birzu:fluctuations} and introduce the spatially-resolved heterozygosity

\begin{equation}
H(t, \zeta_1, \zeta_2) = \langle f(t, \zeta_1)[ 1 - f(t, \zeta_2) ] + f(t, \zeta_2)[ 1 - f(t, \zeta_1) ] \rangle,
\label{eq:H_definition}
\end{equation}

\noindent where~$\langle \cdot \rangle$ denotes the average over independent realizations. As defined,~$H(t, \zeta_1, \zeta_2)$ is the probability of sampling different genotypes from location~$\zeta_1$ and~$\zeta_2$ in the co-moving reference frame. 
We use the shorter term heterozygosity throughout to refer to~$H(t,\zeta_1,\zeta_2)$ when it is unlikely to cause confusion.

To solve for~$H$, we assume the fluctuations are small enough that we can approximate the front profile by the deterministic solution given by Eq.~\eqref{eq:deterministic_n}, i.e.~$\rho(t, \zeta) \approx \rho_d(\zeta)$. Taking the time derivative of~\eqref{eq:H_definition} and substituting~$\pd{f}{t}$ from \eqref{eq:fi_stochastic} we obtain

\begin{equation}
\label{eq:dH_dt}
\pd{H}{t} = [\mathcal{L}_{\zeta_1} + \mathcal{L}_{\zeta_2}]H - \frac{\gamma_{f}(\rho_d)}{N\rho_d} \delta (\zeta_1 - \zeta_2) H,
\end{equation}

\noindent where

\begin{equation}
\mathcal{L}_{\zeta} = \pd{}{\zeta} \left[ D(\rho_d)\pd{}{\zeta} \right] + \left[ v_d + 2D(\rho_d) \pd{\ln \rho_d}{\zeta} \right] \pd{}{\zeta}.
\label{eq:L_definition}
\end{equation}

The equation for~$H$ is a linear differential equation, which can be solved by eigenvector decomposition. At long times, the solution is described by the eigenvector with the largest eigenvalue, which we identify with~$-1/N_e$, following Eq. (8). In general, it is difficult to find the eigenvalue of the full problem, but in the limit of large~$N$, the last term in Eq.~\eqref{eq:dH_dt} can be treated as a perturbation. Using standard techniques, we obtain the following equation for the eigenvalue:

\begin{equation}
\frac{1}{N_e} = \frac{1}{N} \frac{ \int d\zeta_1 d\zeta_2 \delta (\zeta_1 - \zeta_2) \frac{\gamma_{f}[\rho_d(\zeta_1)]}{\rho_d(\zeta_1)} L(\zeta_1, \zeta_2) R(\zeta_1, \zeta_2) }{\int d\zeta_1 d\zeta_2 L(\zeta_1, \zeta_2)R(\zeta_1, \zeta_2)},
\label{eq:Ne_formal}
\end{equation}

\noindent where~$L(\zeta_1, \zeta_2)$ and~$R(\zeta_1, \zeta_2)$ are the left and right eigenvectors of the operator~$\mathcal{L}_{\zeta_1} + \mathcal{L}_{\zeta_2}$ with the largest eigenvalues.

\subsection{Determining the left and right eigenvectors}

The expression derived previously in Eq.~\eqref{eq:Ne_formal} represents a formal solution for~$N_e$. In order to compute a solution we need to find~$R(\zeta_1,\zeta_2)$ and~$L(\zeta_1,\zeta_2)$. Note that we must use only the negative spectrum of the operator~$\mathcal{L}_{\zeta_1} + \mathcal{L}_{\zeta_2}$, since~$H$ is bounded. The spectrum, therefore, has an upper bound at zero. It is relatively easy to show by inspection that an eigenvector with zero eigenvalue exists, so we use this as a starting point of our analysis.

We begin with the right eigenvector, which is the solution to the equation

\begin{equation}
\label{eq:r_formal_eq}
[\mathcal{L}_{\zeta_1} + \mathcal{L}_{\zeta_2}]R(\zeta_1,\zeta_2) = 0,
\end{equation}

which can be solved immediately:

\begin{equation}
R(\zeta_1, \zeta_2) = 1.
\label{eq:R_solution}
\end{equation}

From Eq.~\eqref{eq:R_solution} we can derive~$L(\zeta_1, \zeta_2)$ by standard methods. This derivation can be found in standard textbooks on the Fokker-Planck equation, such as Ref.~\cite{risken:fpe} (see also Sec. V in the SI of Ref.~\cite{birzu:fluctuations} for an analogous derivation for constant ~$D$). Here, we skip these technical details and give the final result:

\begin{equation}
L(\zeta_1, \zeta_2) = \rho_d^2(\zeta_1)\rho_d^2(\zeta_2)e^{v_d \int \frac{d\zeta_1}{D(\rho_d(\zeta_1))}}e^{v_d \int \frac{d\zeta_2}{D(\rho_d(\zeta_2))}}.
\label{eq:L_solution}
\end{equation}
 
Upon substituting~\eqref{eq:L_solution} and~\eqref{eq:R_solution} into~\eqref{eq:Ne_formal} we obtain

\begin{equation}
N_e = N \frac{ \left\{ \int_{-\infty}^{\infty} [\rho_d(\zeta)]^2 \exp \left[ v_d\int_{\mu}^{\zeta} \frac{d\xi}{D(\rho_d(\xi))} \right]d\zeta \right\}^2 }{ \int_{-\infty}^{\infty} \gamma_f(\rho_d) [\rho_d(\zeta)]^3 \exp \left[ 2v_d\int_{\mu}^{\zeta} \frac{d\xi}{D(\rho_d(\xi))} \right]d\zeta }.
\label{eq:Ne}
\end{equation}

In the expressions above, we explicitly account for the integration constant from the integral over~$[D(\rho_d)]^{-1}$ by introducing the lower integration limit~$\mu$. Note that the contribution from the lower limit in the numerator is cancelled by the denominator and the expression is independent of~$\mu$.

\subsection{Scaling of~$N_e$ in pulled, semi-pushed, and fully-pushed expansions}

Clearly, if the integrals in Eq.~\eqref{eq:Ne} converge then~$N_e \sim N$. As we now show, a different scaling appears if one of the integrals is divergent. This can be seen by considering the integral in the denominator. All factors in the integrand saturate at constant values or go to zero in the limit of~$\zeta \rightarrow -\infty$. Therefore, if the integral fails to converge it can only do so in the limit~$\zeta \rightarrow +\infty$. In this limit,~$\rho_d(\zeta)$ decays exponentially (see Eq.~\eqref{eq:solution_front}). It is convenient to express the decay rate using the normalized velocity~$\nu$:

\begin{equation}
\label{eq:k_linear}
k = \frac{v_d}{2D(0)}\left( 1 + \sqrt{1 - \frac{1}{\nu^2}} \right).
\end{equation}

Introducing the exponential solution from above into Eq.~\eqref{eq:Ne} and keeping only the dominant factors gives the asymptotic behavior of the effective population size:

\begin{equation}
\label{eq:asympt_Ne}
N_e \propto N \left[ \int^{+\infty} e^{\frac{v_d\zeta}{2D(0)} \left( 1 - 3\sqrt{1 - \frac{1}{\nu^2}} \right)} d\zeta \right]^{-1}.
\end{equation}

Thus, the convergence of the integral above is determined by the sign of~$1 - 3\sqrt{1 - \frac{1}{\nu^2}}$. The value of the normalized velocity at which the linear scaling breaks down then follows immediately:

\begin{equation}
\label{eq:nu_critical}
\nu_c = \frac{3}{2\sqrt{2}}.
\end{equation}

The divergence identified in Eq.~\eqref{eq:asympt_Ne} is due to the region of the front where~$\rho_d(\zeta) \ll 1$. In actual populations, this region is finite and no divergence occurs. 
We can account for this in our continuum theory by imposing a cutoff on the population density. 
An naive choice for the cutoff is~$\zeta_c$ such that~$\rho_d(\zeta_c) = \frac{1}{Na}$, where~$a$ is some length scale comparable to the size of a deme. Note that~$\zeta_c$ is the position of the last unoccupied deme.
However, this naive cutoff fails to account for the large fluctuations in the density at the edge of the expansion.
The correct expressions were computed in Refs.~\cite{brunet:phenomenological_pulled, birzu:fluctuations} and are given by

\begin{subequations} \label{eq:cutoff}
\begin{numcases}{\zeta_c =}
 \frac{1}{k} \left[ \ln (Nw) + 3\ln \ln (Nw) \right], \quad\quad \nu = 1,\label{eq:pulled_cutoff}\\
\frac{1}{q}\ln (Nw), \quad\quad \nu > 1\label{eq:pushed_cutoff},
\end{numcases}
\end{subequations}

\noindent where~$w\sim 1/k$ is the effective width of the front and~$q$ is given by Eq.~\eqref{eq:k_q_front}.

Substituting Eqs.~\eqref{eq:cutoff} into~\eqref{eq:Ne} gives the formula for the effective population size from Eq. (12) in the main text, with the following values for the constants~$C_1$,~$C_2$, and~$C_3$:

\begin{equation}
\label{eq:ne_constants}
\begin{aligned}
C_1 &= \frac{1}{2\pi^2 r_0},\\
C_2 &= \frac{\gamma_f(0) \exp{ \left[ 2v_d \left( \int_{\mu}^{\zeta_c} \frac{d\xi}{D(\rho_d(\xi))} - \frac{\zeta_c}{D(0)} \right) \right]}}{\left[ \int_{-\infty}^{\infty}\rho_d^2 \exp \left[ v_d \int_{\mu}^{\zeta_c} \frac{d\xi}{D(\rho_d(\xi))} \right]d\zeta \right]^2},\\
C_{3} &= \frac{\int_{-\infty}^{\infty} \gamma_f(\rho_d)\rho_d^3 \exp \left[ 2v_d \int_{\mu}^{\infty} \frac{d\zeta}{D(\rho_d(\zeta))} \right]d\zeta}{\left\{ \int_{-\infty}^{\infty}\rho_d^2 \exp \left[ v_d \int_{\mu}^{\infty} \frac{d\zeta}{D(\rho_d(\zeta))} \right]d\zeta \right\}^2}.
\end{aligned}
\end{equation}

%**************************************************************************
% Diffusion constant of the front
%**************************************************************************
\section{Front diffusion} \label{sec:Df}

Demographic fluctuations, represented by the stochastic term in Eq.~\eqref{eq:n_stochastic}, lead to a stochastic wandering of the front. In this section, we show that this wandering is diffusive and derive the general formula for the diffusion constant~$D_{f}$. The derivation is an extension of the perturbation theory developed in Refs.~\cite{mikhailov:diffusion} and~\cite{rocco:diffusion, meerson:velocity_fluctuations}. The presentation closely follows the analogous calculation in Ref.~\cite{birzu:fluctuations} (SI, Sec. VI). We would also like to point the reader to a potentially relevant Ref.~\cite{balasuriya:wavespeed_chemotaxis} that used an alternative method to obtain corrections to front velocity.

\subsection{Perturbation theory for demographic fluctuations}

We begin with Eq.~\eqref{eq:n_stochastic} written in a more convenient form:

\begin{equation}
\frac{\partial \rho}{\partial t} = \pd{}{x}\left[ D(\rho)\pd{\rho}{x}\right] + r(\rho)\rho +\frac{1}{\sqrt{N}}\Gamma(\rho)\eta
\label{eq:diffusion_problem}
\end{equation}

\noindent where~$\rho=n/N$ is the normalized population density and~$\Gamma$ denotes the strength of the noise term:

\begin{equation}
\label{eq:Gamma}
\Gamma(\rho) = \sqrt{\gamma_n(\rho)\rho}.
\end{equation}

We seek a solution to Eq.~\eqref{eq:diffusion_problem} of the form

\begin{equation}
\label{eq:perturbed_profile}
\rho(t,x) = \rho_d(x-v_dt-\xi(t)) + \Delta \rho (t, x-v_dt-\xi(t)),
\end{equation}

\noindent where~$v_d$ and~$\rho_d$ are given by the solution of the deterministic equation ($\Gamma=0$). Substituting Eq.~\eqref{eq:perturbed_profile} into Eq.~\eqref{eq:diffusion_problem} results in the equation for~$\Delta \rho$

\begin{equation}
\label{eq:perturbation_first_order_diffusion}
\frac{\partial \Delta \rho}{\partial t} - \mathcal{L}_{f} \Delta \rho =  \rho'_d \xi' + \frac{1}{\sqrt{N}}\Gamma(\rho_d)\eta,
\end{equation}

\noindent where we introduced the operator~$\mathcal{L}_{f}$:

\begin{equation}
\label{eq:cal_L}
\begin{aligned}
\mathcal{L}_{f}  = & \dif{}{\zeta} \left[ D(\rho_d) \dif{}{\zeta} \right] + [ v_d + D'(\rho_d) \rho_d' ]\dif{}{\zeta}\\
	& + \left[ r_d(\rho_d) + r_d'(\rho_d)\rho_d + D'(\rho_d)\rho_d'' + D''(\rho_d)(\rho_d')^2 \right].
\end{aligned}
\end{equation}

To solve~\eqref{eq:perturbation_first_order_diffusion} we use the fact that~$\mathcal{L}_{f}$ has a pair of left and right eigenvectors with zero eigenvalue. 
We denote them by~$L_{f}$ and~$R_{f}$, respectively.
We shown below that the existence of these eigenvectors is a consequence of the translational symmetry in Eq.~\eqref{eq:deterministic_n}.

Multiplying Eq.~\eqref{eq:perturbation_first_order_diffusion} by~$L_{f}(\zeta)$ and integrating over~$\zeta$ gives

\begin{equation}
\label{eq:projection_on_L}
\pd{}{t}\int_{-\infty}^{+\infty} L_{f} \Delta\rho d\zeta  =  \int_{-\infty}^{+\infty} L_{f} \rho'_d\xi' d\zeta +  \frac{1}{\sqrt{N}}\int_{-\infty}^{+\infty} L_{f} \Gamma(\rho_d)\eta d\zeta.
\end{equation}

\noindent We now use the fact that $\int_{-\infty}^{+\infty} L_{f} (\zeta) \Delta \rho(\zeta) d\zeta=0$. The projection of~$\Delta \rho$ on~$L_{f}$ vanishes because we separated translations of~$\rho_d$ from shape fluctuations in Eq.~\eqref{eq:perturbed_profile} are excluded from the fluctuations of the front shape and are instead included through~$\xi(t)$. Otherwise, according to Eq.~(\ref{eq:projection_on_L}), $\int_{-\infty}^{\infty} L_{f} \Delta\rho d\zeta$~would perform an unconstrained random walk and grow arbitrarily large, which would violate the assumption that~$\Delta \rho$ is small. After imposing~$\int_{-\infty}^{+\infty} L_{f} \Delta\rho d\zeta=0$, we obtain

\begin{equation}
\label{eq:xi_prime}
\xi'(t) = -\frac{1}{\sqrt{N}} \frac{\int_{-\infty}^{+\infty}L_{f}(\zeta)\Gamma(\rho_d(\zeta))\eta(t,\zeta) d\zeta}{\int_{-\infty}^{+\infty}L_{f}(\zeta)\rho'_d(\zeta)d\zeta}.
\end{equation}

From Eq.~\eqref{eq:xi_prime} it is immediately clear that

\begin{equation}
\langle \xi'(t) \rangle = 0.
\end{equation}

\noindent Therefore, the translational motion of the front is stochastic and unbiased. To characterize this motion we determine the diffusion constant of the front~$D_{f}$ defined by

\begin{equation}
\label{eq:definition_Df}
D_{f} = \frac{\mathrm{Var} \{X_{f}^2\}}{2t} = \frac{\langle \xi^2\rangle}{2t},
\end{equation}

\noindent where~$X_{f}$ is the position if the front. Substituting Eq.~\eqref{eq:xi_prime} into Eq.~\eqref{eq:definition_Df} gives the formal expression for the front diffusion constant:

\begin{equation}
D_{f} = \frac{1}{2N} \frac{\int_{-\infty}^{\infty} \gamma_n(\rho) [L_{f}(\zeta)]^2 \rho(\zeta) d\zeta }{\left[ \int_{-\infty}^{\infty} L_{f}(\zeta) \rho'(\zeta) d\zeta \right]^2}.
\label{eq:Df_w_l}
\end{equation}

\subsection{The left eigenvector}

To find~$L_{f}$ we first find~$R_{f}$ and then obtain~$L_{f}$ by transforming operator~$\mathcal{L}_{f}$ into Hermitian form.
It is easy to check that~$R_{f}(\zeta) = \rho_d'(\zeta)$.
This can be seen from the following heuristic argument. Since the unperturbed problem is translationally invariant, any shift in the deterministic front solution~$\rho_d(\zeta)$ is also a solution. In particular, for an infinitesimal shift we have~$\rho_d(\zeta + \delta \zeta) \approx \rho_d(\zeta) + \rho_d'(\zeta) \delta \zeta$. Thus, we expect a perturbation~$\Delta \rho \propto \rho_d'(\zeta)$ should leave the left hand side of Eq.~\eqref{eq:perturbation_first_order_diffusion} unchanged. This can only happen if $R_{f}(\zeta) \propto \rho_d'(\zeta)$.

We now proceed to find the corresponding left eigenvector. This can be done by finding a function~$\beta(\zeta)$ such that

\begin{equation}
\label{eq:H_transformation_2}
\mathcal{L}_{f} = e^{\beta(\zeta)} \mathcal{H}_{f} e^{-\beta(\zeta)},
\end{equation}

\noindent and~$\mathcal{H}_{f}$ is a Hermitian operator.
Because~$\mathcal{H}_{f}$ is Hermitian, the left and right eigenvectors coincide and can be denoted by~$\psi$. Then it can be easily shown that

\begin{equation}
\label{eq:psi_transform}
\begin{aligned}
L_{f}(\zeta) = e^{-\beta(\zeta)} \psi(\zeta),\\
R_{f}(\zeta) = e^{\beta(\zeta)} \psi(\zeta),
\end{aligned}
\end{equation}

and

\begin{equation}
\label{eq:L_R_transform}
L_{f}(\zeta) = e^{-2\beta(\zeta)}\rho_d'(\zeta).
\end{equation}

Substituting~$\psi(\zeta) = e^{-\beta(\zeta)}\rho_d'(\zeta)$ into the eigenvalue problem~$\mathcal{H}_{f} \psi = 0$ gives the expression for~$\beta$:

\begin{equation}
\beta(\zeta) = - \frac{v_d}{2}\int_{\mu}^{\zeta} \frac{d\xi} {D[n_d(\xi)]}   - \frac{1}{2} \ln[D[n_d(\zeta)]].
\label{eq:beta_solution}
\end{equation}

Finally, upon substituting~\eqref{eq:beta_solution} and~\eqref{eq:L_R_transform} into Eq.~\eqref{eq:Df_w_l} we arrive at the final expression for the diffusion constant

\begin{equation}
D_{f} = \frac{1}{2N} \frac{\int_{-\infty}^{\infty} \gamma_n(\rho_d) D^2(\rho_d(\zeta)) \rho_d(\zeta) [\rho_d'(\zeta)]^2 \exp \left[ 2v_d\int_{\mu}^{\zeta} \frac{d\xi} {D(\rho_d(\xi))}  \right] d \zeta }{\left\{ \int_{-\infty}^{\infty} D(\rho_d(\zeta)) [\rho_d'(\zeta)]^2 \ \exp \left[ v_d\int_{\mu}^{\zeta} \frac{d\xi} {D(\rho_d(\xi))}  \right] d \zeta \right\}^2}.
\label{eq:Df}
\end{equation}

As we have already shown,~$\rho_d(\zeta) \propto e^{-k\zeta}$ when~$\zeta \rightarrow +\infty$. Therefore,~$\rho_d'(\zeta) \propto e^{-k\zeta}$ as well and the analysis of the integrals in Eq.~\eqref{eq:Df} is completely analogous to the one for~$N_e^{-1}$, and the same scaling regimes apply.

%**************************************************************************
% Exactly solvable models
%**************************************************************************
\section{Exactly solvable models} \label{sec:exact_models}

Given importance of exact solutions, we present a few exactly solvable models of range expansions with density-dependent dispersal. Throughout this section we assume~$\gamma_n(\rho)=\gamma_n^0 (1 - \rho)$ and~$\gamma_f(\rho)=\gamma_f^0$, which provide the correct continuum description of our simulations (see Sec.~\ref{sec:simulations}). Note that exact solutions can also be obtained for other forms of gammas (see Secs. VI and VIII of SI in Ref.~\cite{birzu:fluctuations})

We begin with the model presented in the main main text and defined by the following equations:

\begin{equation}
\begin{aligned}
D(\rho) &= D_0 \left( 1 +  A\rho \right),\\
r(\rho) &= r_0 (1 - \rho).
\end{aligned}
\end{equation}

Recently, exact solutions for the expansion velocity and steady-state profile for this model were found~\cite{kawasaki:long_range_taxis}. The exact solution shows that the wave is pulled for~$A<2$ and pushed for~$A > 2$. In the pulled regime, the solution is given by

\begin{equation}
\label{eq:v_linear}
\begin{aligned}
v &=  2 \sqrt{r_0D_0}, \\
\rho(\zeta) &\sim e^{\sqrt{\frac{r_0}{D_0}}\zeta},
\end{aligned}
\end{equation}

and in the pushed regime, the solution is given by

\begin{equation}
\label{eq:v_linear}
\begin{aligned}
v &= \sqrt{r_0 D_0} \left( \sqrt{A/2} + \sqrt{2/A} \right), \\
\frac{(1 - \rho)^{1 + A}}{\rho} &= e^{\sqrt{\frac{r_0A}{2D_0}}\zeta}.
\end{aligned}
\end{equation}

To the best of our knowledge, this is the only model in which a transition from a pulled to a pushed wave due to density-dependent dispersal can be demonstrated analytically. Using these results we can compute the effective population size and the front diffusion constant exactly in the fully-pushed regime. They are given by the following expressions:

\begin{equation}
\label{eq:Ne_linear_model}
N_e = \frac{6\pi N}{\gamma_f^0} \sqrt{\frac{2D_0}{r_0A}} \frac{A^2}{(A + 2)(A + 4)} \cot{\frac{2\pi}{A}},
\end{equation}

and

\begin{equation}
\label{eq:Df_linear_model}
D_f = \frac{\gamma_n^0}{10\pi N} \sqrt{\frac{2D_0}{r_0 A}} \frac{A (A + 4) (3A + 4)}{A + 1} \tan{\frac{2\pi}{A}}.
\end{equation}

A separate class of exactly solvable models are given in Ref.~\cite{petrovskii:exact_models} and are defined by

\begin{equation}
\label{eq:exactly_solvable}
\begin{aligned}
D(\rho) &= D_0 \rho^q,\\ 
r(\rho) &= r_0\rho^p \left( 1 - \rho^{k} \right),
\end{aligned}
\end{equation}

\noindent with

\begin{equation}
\label{eq:exactly_solvable_condition}
k = p + q - 1.
\end{equation}

In this case, since~$D(0) = 0$, the waves are always fully-pushed. The exact solutions for the velocity and profile are given by the following formulas:

\begin{equation}
\begin{aligned}
v &= \frac{zD_0}{\lambda_q}\\
\rho(\zeta) &=
\begin{cases}
\left( 1 - e^{ \frac{\zeta}{\lambda_q}} \right)^{1/q}, & \zeta < 0\\
0, & \zeta \ge 0,
\end{cases}
\end{aligned}
\label{eq:exact_solution_pq}
\end{equation}

\noindent where we used 

\begin{equation}
\label{eq:lambda_q}
\begin{aligned}
\lambda_q &= \sqrt{\frac{(q + 1)D_0}{r_0}},\\
z &= \sqrt{\frac{q + 1}{p + q}}.
\end{aligned}
\end{equation}

In this case, the results for the effective population size and front diffusion constant are 

\begin{equation}
N_e = \frac{\lambda_q N}{\gamma_f^0} \frac{ \left[ \mathrm{B} \left( \frac{q + 2}{q} -z, z \right) \right]^2 }{ \mathrm{B} \left( \frac{q + 3}{q} - 2z, 2z \right) }.\label{eq:Lambda_exact}
\end{equation}

\noindent and

\begin{equation}
D_f = \frac{q^2 \gamma_n^0 \lambda_q}{2N} \frac{ \mathrm{B} \left( \frac{q + 3}{q} - 2z, 2 + 2z \right) - \mathrm{B} \left( \frac{q + 4}{q} - 2z, 2 + 2z \right)}{\left[ \mathrm{B} \left( \frac{2 - qz}{q}, 2 + z \right) \right]^2}.
\end{equation}

%**************************************************************************
% Simulations and data analysis
%**************************************************************************
\section{Computer simulations} \label{sec:simulations}

In this section, we explain the details of our computer simulations and the subsequent data analysis.

We simulated range expansions of two neutral genotypes in a one-dimensional habitat modeled by an array of patches separated by distance~$\Delta x$; the time was discretized in steps of duration~$\Delta t$. Thus, the abundance of each genotype was represented as~$n_i(t,x)$, where~$i\in\{1,2\}$ is the index of the genotype, and~$t$ and~$x$ are integer multiples of~$\Delta t$ and~$\Delta x$. Each time step, we updated the abundance of both genotypes simultaneously by drawing from a multinomial distribution with~$N$ trials and probability~$p_i$ to sample genotype~$i$. The values of~$p_i$ were chosen such that~$p_i N$ were equal to the expected abundances of the genotypes following dispersal and growth:

\begin{equation}
p_i = \frac{ 1 +  r( \tilde{n} )\Delta t }{N} \left[ n_i(t) - m(n(t, x)) n_i(t, x) + \frac{m(n(t, x - \Delta x))}{2}n_i(t, x - \Delta x) + \frac{m(n(t, x + \Delta x))}{2}n_i(t, x + \Delta x) \right],
\label{eq:p_i}
\end{equation}
 
\noindent where

\begin{equation}
\label{eq:tilde_n}
\begin{aligned}
\tilde{n} =& [1-m(n(t,x))]n_1(t,x) + \frac{m(n(t,x-\Delta x)}{2}n_1(t,x-\Delta x) + \frac{m(n(t,x+\Delta x))}{2}n_1(t,x+\Delta x) \\
& + [1-m(n(t,x))]n_2(t,x) + \frac{m(n(t,x-\Delta x)}{2}n_2(t,x-\Delta x) + \frac{m(n(t,x+\Delta x))}{2}n_2(t,x+\Delta x),
\end{aligned}
\end{equation}

which is the expected total population density after dispersal, but before growth. For simplicity, we set both~$\Delta x$ and~$\Delta t$ to~$1$ in all simulations.

\subsection{Mapping simulation parameters to continuum model}

To find the reaction-diffusion equation corresponding to our simulation algorithm we perform a Taylor expansion of the population density:

\begin{equation}
\begin{aligned}
n_i (t+\Delta t, x) &= n_i (t, x) + \Delta t\pd{n_i}{t} \\
n_i(t, x \pm \Delta x) &= n_i(t, x) \pm \Delta x\pd{n_i}{x}  + \frac{\Delta x^{2}}{2} \pdd{n_i}{x} \\
m \left[ n(t, x \pm \Delta x) \right] &= m[n(t, x)] + m'[n(t, x)] \left( \pm \Delta x \pd{n}{x} + \frac{\Delta x^{2} }{2} \pdd{n}{x} \right) + \frac{\Delta x^{2}}{2} m''[n(t, x)] \left( \pd{n}{x} \right)^2,
\end{aligned}
\label{eq:Taylor_expansion}
\end{equation}
\noindent where we've kept terms up to first order in~$\Delta t$ and second order in~$\Delta x$.

Substituting~\eqref{eq:Taylor_expansion} into Eq.~\eqref{eq:p_i} and summing over the two genotypes gives the equation for the total population density:

\begin{equation}
\begin{aligned}
\pd{n}{t} & = \frac{\Delta x^{2}}{2\Delta t} \left[ m(n)   + m'(n) n \right] \pdd{n}{x} + \frac{\Delta x^{2}}{2\Delta t} \left[ m''(n) n + 2m'(n) \right] \left(\pd{n}{x} \right)^{2}  + r ( n ) n,
\end{aligned}
\label{eq:discrete_intermediate}
\end{equation}

\noindent which can be rewritten as

\begin{equation}
\pd{n}{t}  = \pdd{}{x}\left[ \frac{\Delta x^2 m(n)}{2\Delta t} n \right] + r (n) n.
\label{eq:discrete_pde}
\end{equation}

In the continuum limit, when~$r(n)\Delta t\ll1$ and~$k\Delta x\ll1$, our model becomes equivalent to Eq.~\eqref{eq:n_stochastic} for the population density and to Eq.~\eqref{eq:fi_stochastic} for the relative fraction of the two genotypes with

\begin{equation}
\begin{aligned}
\tilde{D}(n) &= \frac{\Delta x^2 m(n)}{2\Delta t}, \\
\gamma_n (n) &= \frac{1-n/N}{\Delta t}, \\
\gamma_f (n) &= \frac{1}{\Delta x\Delta t}.
\end{aligned}
\end{equation}

Note that~$D(n) = \tilde{D}(n) + \tilde{D}'(n)n$ as discussed in Sec.~\ref{sec:dispersal}.

\subsection{Boundary and initial conditions}
The most direct approach to simulating a range expansion is to use a stationary habitat, in which the range expansion proceeds from one end to the other. This approach is, however, inefficient because the computational time grows quadratically with the duration of the simulation. Instead, we took advantage of the fact that all population dynamics are localized near the expansion front and simulated only a region of 300 patches comoving with the expansion. Specifically, every simulation time step, we shifted the front backward if the total population inside the simulation array~$n_{\mathrm{array}}$ exceeded~$150N$, i.e. half of the maximally possible population size. The magnitude of the shift was equal to~$\lfloor (n_{\mathrm{array}}-150N)/N \rfloor+1$. The population density in the patches that were added ahead of the front was set to zero, and the number of the individuals moved outside the box was stored, so that we could compute the total number of individuals~$n_{\mathrm{tot}}$ in the entire population including both inside and outside of the simulation array.

Our choice of~$300$ patches in the simulation array was sufficient to ensure that at least one patch remained always unoccupied ahead of the expansion front and that the patches shifted outside the array were always at the carrying capacity. While neglecting the effect of patches deep inside the bulk is an approximation, its effect is relatively small if the size of the box is large compared to the front width. This can be seen from Eq.\eqref{eq:Ne}, where the contribution of the bulk to the two integrals decays exponentially as~$~e^{-k_b | \zeta |}$. In our case,~$1/k_b$ is of the order of~$10$ patches and, thus, the error in determining~$N_e$ is negligible.

We initialized all simulations by leaving the right half of the array unoccupied and filling the left half to the carrying capacity. In each occupied patch, we determined the relative abundance of the two neutral genotypes by sampling from the binomial distribution with~$N$ trials and equal probabilities of choosing each of the genotypes.

\subsection{Computing front velocity}
The velocity of the front was measured by fitting~$n_{\mathrm{tot}}/N$ to~$vt+\mathrm{const}$. For this fit, we discarded the first~$10\%$ of the total simulation time (1000 generations for the shortest runs) 
to account for the transient dynamics. The length of the transient is the largest for pulled waves and is specified by the following result from Ref.~\cite{saarloos:review}:

\begin{equation}
v(t) = v_{\mathrm{\textsc{F}}} \left[ 1 - \frac{3}{4r_0 t} + \mathcal{O}\left( t^{-3/2} \right) \right].
\end{equation}

\noindent Thus, discarding time points prior to~$t\sim1/r_0$ was sufficient to eliminate the transient dynamics in all of our simulations.

\subsection{Computing heterozygosity and the rate of its decay}\label{DA:diversity}
For each time point, the heterozygosity~$h$ was computed as follows

\begin{equation}
\label{eq:h_definition}
h(t) = \frac{1}{n_{\mathrm{nonzero}}}\sum_{x, n(t, x) \neq 0}\frac{2n_1(t,x)n_2(t,x)}{[n_1(t,x)+n_2(t,x)]^2},
\end{equation} 

\noindent where~$n_{\mathrm{nonzero}}$ is the number of occupied patches. The average heterozygosity~$H$ was then obtained by averaging over independent simulation runs. To compute~$N_e$, we fitted~$\ln H$ to~$-t/N_e +\mathrm{const}$.

Because of the initial transient dynamics and large uncertainty in~$H$ for large~$t$, due to the small number of replicas with~$H > 0$, we restricted the analysis of~$H(t)$ to~$t\in(t_{\mathrm{i}},t_{f})$.
The value of~$t_{f}$ was chosen such that at least~$50$ simulations had non-zero heterozygosity at~$t=t_{f}$. The value of~$t_{\mathrm{i}}$ was chosen to maximize the goodness of fit~($R^2$) between the fit to~$H\sim e^{- t/N_e}$ and the data subject to the constraint that~$t_{f}-t_{\mathrm{i}}>1000$. The latter constraint ensured that we had a sufficient number of uncorrelated data points for carrying out the fitting procedure to determine~$N_e$.

\subsection{Comparison between simulations and theory}

Table S1 shows a comparison between the theoretical prediction and the exponents obtained by fitting the data in Fig. 4 (main text). In the fully-pushed regime, the data are consistent with the linear scaling predicted by the theory. In the semi-pushed regime, the agreement is less good for the square root density-dependence and, to a lesser extent, for the linear density-dependence. This may reflect a larger contribution of the subleading corrections to~$N_e$ in these models. For pulled waves, the asymptotic~$\log^3 N$ scaling sets in at very large values of~$N$, as demonstrated in Ref.~\cite{brunet:phenomenological_pulled}. 
Nevertheless, as Figs. 4 and 5 show, our simulations are consistent with~$N_e \sim \log^3 N$, as predicted by theory.

\begin{table}[h!]
\label{t:comparison}
\begin{center}
\begin{tabular}{ | c c c c c | }
\hline
expansion class & $\nu$  & D(n) & Theory exponent & Best fit exponent \\
\hline
semi-pushed & $1.03$ & $D_0(1 + A_{1/2} \sqrt{n/N})$ & $0.63$ & $0.55 \pm 0.01$ \\
semi-pushed & $1.03$ & $D_0(1 + A_{1} n/N)$ & $0.63$ & $0.58 \pm 0.04$ \\
semi-pushed & $1.03$ & $D_0(1 + A_{2} n^2/N^2)$ & $0.63$ & $0.65 \pm 0.04$ \\
fully-pushed & $1.5$ & $D_0(1 + A_{1/2} \sqrt{n/N})$ & $1$ & $0.98 \pm 0.03$ \\
fully-pushed & $1.5$ & $D_0(1 + A_{1} n/N)$ & $1$ & $1.01 \pm 0.04$ \\
fully-pushed & $1.5$ & $D_0(1 + A_{2} n^2/N^2)$  & $1$ & $0.98 \pm 0.04$ \\
\hline
\end{tabular}
\caption{\textbf{Comparison between theoretical predictions and best fit for different models of dispersal.} The best fit exponents were obtained by preforming a least squares fit of the data to the power law~$y(x)=ax^b$, using the~\texttt{curve\_fit} function from the scipy package (version 1.1; \cite{jones:scipy}). The errors in the exponents represent two standard deviations.}
\end{center}
\end{table}

\printbibliography

\end{document}